\documentclass{ws-ijmpc}

\begin{document}

\markboth{Nuno Crokidakis}
{A simple mechanism leading to first-order phase transitions in a model of tax evasion}

\catchline{}{}{}{}{}

\title{A simple mechanism leading to first-order phase transitions in a model of tax evasion}

\author{Nuno Crokidakis $^{*}$}

\address{
Instituto de F\'{\i}sica, \hspace{1mm} Universidade Federal Fluminense \\
 Niter\'oi - Rio de Janeiro, \hspace{1mm} Brazil \\ 
$^{*}$ nuno@mail.if.uff.br}

\maketitle

\begin{history}
\received{Day Month Year}
\revised{Day Month Year}
\end{history}

\begin{abstract}
\noindent
In this work we study a dynamics of tax evasion. We considered a fully-connected population divided in three compartments, namely honest tax payers, tax evaders and susceptibles, a class that is composed by honest tax payers that can become evaders. We consider a contagion model where the transitions among the compartments are governed by probabilities. Such probabilities represent the possible interactions among the indiviudals, as well as the government's fiscalization. We show by analytical and numerical calculations that the emergence of tax evaders in the population is associated with an active-absorbing nonequilibrium first-order phase transition. In the absorbing phase only honest tax payers survive in the steady states of the model, and we observe a coexistence of the three subpopulations in the active phase.


\end{abstract}

\ccode{PACS Nos.: 05.10.-a, 05.70.Jk, 87.23.Ge, 89.75.Fb}

\section{Introduction}

\qquad Socioeconomic problems have been recently the foccus of sociophysics researchers \cite{econ_book,pmco_book}. The basic approach is usually dominated by agent-based models, which allow us to understand the emergente collective phenomena of such systems. Among the studied problems, one of great interest is tax evasion dynamics, which is interesting from the practical point of view because tax evasion remains to be a major predicament facing governments \cite{bloom,prinz,andreoni}.

Economists studied models of tax evasion during several years \cite{gachter,frey,follmer,slemrod,davis,wenzel,hood,salmina1}, and more recently physicists also became interested in the subject \cite{zaklan,lima1,lima2,llacer,seibold1,meu_econo,bertotti} (for recent reviews, see \cite{bloom,prinz,hokamp,seibold2}). The authors in \cite{salmina1} discussed that the management of tax arrears it's not only the performance of the tax authorities of their duties to collect the debt, but  also  the  formation  of  the  system  of  economic  relations  that  arise  between  the  state  (local  government),  business  entities  and  individuals  regarding  the  payment  of  fees. Regarding the Russian federation, a recent work verified that although the number of the on-site  tax  inspections  is  currently  being  reduced,  it  does  not  affect  their  performance  efficiency  as  the  tax  control authorities select the auditees more thoroughly \cite{salmina2}. Microscopic models were recently developed. The authors in \cite{bertotti,bertotti2} considered a system of nonlinear differential equations of the kinetic discretized Boltzmann type involving transition probabilities, and combined a tax system with a redistribution process. They showed that a inequality index (the Gini index) of the total population increases when the evasion level is higher, but does not depend significantly on the evasion spread.

Physics models based on the Ising model were proposed to analyze tax evasion dynamics \cite{zaklan,lima1,lima2}. These works analyze how enforcement rules, as well as agent-agent interactions can be combined to reduce tax evasion and tax evasion fluctuations. As an extension of such works, a recent paper proposed a diluted Ising model with competing interactions to model tax evasion dynamics \cite{diluted}. The authors considered social variables such as the audit period and its effects over the percentage of evasion, in order to analyze the behavior of tax evasion in Colombia. They found that the magnetic field, representing general government policies, as well as the audit probability, must be complementary if a reduction in tax evasion is desired. Local temperatures and local magnetic fields were also considered in another work \cite{seibold1}.

The author in \cite{meu_econo} studied a more general spin-like model, considering a three-state model. Based on a kinetic exchange opinion model \cite{biswas} together with the enforcement rules of the Zaklan model, it was found that below the critical point of the opinion dynamics the compliance is high, and the punishment rules have a small effect in the population. On the other hand, above the critical point of the opinion dynamics the tax evasion can be considerably reduced by the enforcement mechanism \cite{meu_econo}. Considering a contagion epidemic-like model, a recent work \cite{rafael} showed that the emergence of tax evaders in the population can be associated with an active-absorbing nonequilibrium continuous phase transition.

Rui Barbosa, a Brazilian diplomat, writer, jurist, and politician, said \textit{``To see triumph the nullities, see prosper the dishonor, see grow the injustice, to see agglomerate powers in the hands of the wicked, the man comes to discourage the virtue, laughing at the honor, ashamed to be honest''} \cite{rui}. This is specially true in countries like Brazil, where there is a weak fiscalization and/or light punishment \cite{utsumi,book}. This implies that social pressure of individuals' contacts play an important role in the propagation of social norms, as discussed in recent works \cite{laguna1,laguna2}.

In this work we extend a recent three-state model to analyze tax evasion dynamics. For this purpose, we considered mechanisms of social pressure and enforcement regime. Opinion dynamics models were considered recently to study the addoption of right/wrong behavior in societies \cite{laguna1,laguna2}. Here we adopt a distinct approach, considering an epidemic-like model, where the transition among the classes or compartments are ruled by probabilities \cite{bailey,silvio}, to theoretically study the specific problem of tax evasion. We will see that the emergence of tax evaders in the population can be associated with a first-order nonequilibrium phase transition. The emergence of phase transitons in not usual in models of tax evasion \cite{zaklan,lima1,lima2,llacer,seibold1,meu_econo,bertotti2,diluted,rafael}. In addition, to the best of our knowledge it is the first time that a first-order transition is observed in such models.

This work is organized as follows. In section 2 we define the model's rules and the individuals presented in the population. After, in section 3 we discuss our analytical and numerical results. Finally, in section 4 we present our conclusions and final remarks.


\section{Model}

\qquad We considered a population of $N$ agents. Each individual $i$ ($i=1,2,...,N$) can be in one of three possible states at a given time step $t$, represented by $X_{1}(t)$, $X_{2}(t)$ and $X_{3}(t)$. In other words, $X_{j}$ represents the number of individuals in a given state, with $j=1,2,3$. The state $X_{1}$ represents a \textit{honest tax payer}, i.e., an individual 100$\%$ convinced of his/her honesty, who does not consider evasion. He/she is either habitually compliant or he/she is a recent evader who has become honest as a result of enforcement efforts or social norms. On the other hand, the state $X_{3}$ represents a cheater, i.e, an individual who is an \textit{evading tax payer}. Whether a tax payer continues to evade depends on both enforcement and the effect of social interactions. Finally, the third state $X_{2}$ consists of taxpayers who are dissatisfied with the tax system (perhaps as a result of seeing others evade without being punished). These taxpayers are not actively evading, but they might if the perceived benefits of doing so exceed the perceived costs. For this group, evasion is an option, and so we classify them as \textit{susceptibles}, i.e., they are susceptible to become evaders \cite{davis,meu_econo,rafael}.

Following the model studied in \cite{rafael}, we consider an extra social interaction, namely interactions among honest tax payers $X_1$ and susceptible individuals $X_2$. Due to social pressure of $X_1$ individuals, the susceptible agents $X_2$ hesitate and come back to the $X_1$ compartment. This new transition occurrs with probability $\epsilon$. In addition, we keep the previous two social interactions and the enforcement regime \cite{rafael}. The possible transitions are as follows:
\begin{eqnarray} \label{eq1}
X_{1} + X_{3} & \stackrel{\lambda}{\rightarrow} & X_{2} + X_{3} ~,  \\ \label{eq2}
X_{2} & \stackrel{\alpha}{\rightarrow} & X_{3} ~,  \\  \label{eq_new}
X_{2} + X_{1} & \stackrel{\epsilon}{\rightarrow} & X_{1} + X_{1} ~, \\ \label{eq3}
X_{3} + X_{1} & \stackrel{\delta}{\rightarrow} & X_{1} + X_{1}  ~, \\ \label{eq4}
X_{3} & \stackrel{\beta}{\rightarrow} & X_{1} ~.
\end{eqnarray}

As discussed in \cite{rafael}, Eq. \eqref{eq1} represents an encounter of a honest agent $X_{1}$ with an evader $X_{3}$. In this case, with probability $\lambda$ the honest individual becomes susceptible $X_{2}$. The parameter $\lambda$ can be viewed as the social pressure of evaders over honests. The transition occurrs to the susceptible state, i.e., we consider that the transition from honest to evader is not abrupt, as it is common in tax evasion models \cite{zaklan,lima1,lima2,bertotti,bertotti2,diluted}. The following transition, Eq. \eqref{eq2} represents a spontaneous transition from the susceptible state $X_{2}$ to the evader state $X_{3}$. The enforcement affects the behavior of a susceptible individual through its effect on the perceived costs of evasion (cost-benefit analysis). Thus, we assume that some susceptible tax payers will perceive that the benefits of evasion exceed the costs of evasion in each period, leading these individuals to evade. This is represented by the probability $\alpha$. As discussed above, we consider that the transition from honest to evader is not abrupt: the honest individual first becomes susceptible and after he might become evader.

The new interaction rule is represented by Eq. \eqref{eq_new}. The susceptible agent $X_2$ can be persuaded by a honest one $X_1$ and hesitate to not become an evader $X_3$. In such a case, the susceptible agent returns to the honest compartment $X_1$ with probability $\epsilon$.

Eq. \eqref{eq3} represents the opposite transition in comparison with Eq. \eqref{eq1}. In this case, it represents an encounter of an evader agent $X_{3}$ with a honest tax payer $X_{1}$. In this case, with probability $\delta$ the evader agent becomes honest. The parameter $\delta$ can be viewed as the social pressure of honests over evaders. We can also consider that this last transition occurs to the state $X_{2}$, but for simplicity we consider that the evaders go directly to the honest compartment. 

Finnaly, Eq. (\ref{eq4}) represents another enforcement effect. We consider that
evaders become compliant after they are audited or when their perceptions regarding the costs and benefits of evasion change, either through experience or changing economic conditions \cite{davis}. This transition occurs with probability $\beta$, that can be viewed as a measure of the efficiency of the government's fiscalization. As in the previous case, one can also consider that some evaders might not be rehabilitated when they are audited, remaining susceptible rather than becoming honest, but for simplicity we will not consider those additional transitions.

In the next section we discuss the analytical and numerical results. We will show that the new mechanism represented by the hesitance of susceptible $X_2$ individuals to become evaders $X_3$, under the influence of honest agents $X_1$, leads to the emergence of a discontinuous nonequilibrium phase transition, which is absent for the special case $\epsilon=0$ where only continuous phase transitions were observed \cite{rafael}.


\section{Results}

\qquad In this section we consider the model on a fully-connected graph. Considering the densities of each state, namely $x_{j}=X_{j}/N$ ($j=1,2,3$), we can write master equations for each density as follows:
\begin{eqnarray} \label{eq5}
\frac{d}{dt}\,\,x_{1}(t) & = &  - \lambda\,x_{1}(t)\,x_{3}(t) + \delta\,x_{1}(t)\,x_{3}(t) + \beta\,x_{3}(t) + \epsilon\,x_{1}(t)\,x_{2}(t) ~, \\ \label{eq6}
\frac{d}{dt}\,x_{2}(t) & = & \lambda\,x_{1}(t)\,x_{3}(t)  - \alpha\,x_{2}(t) - \epsilon\,x_{1}(t)\,x_{2}(t)~, \\ \label{eq7}
\frac{d}{dt}\,x_{3}(t) & = & \alpha\,x_{2}(t) - \beta\,x_{3}(t) - \delta\,x_{1}(t)\,x_{3}(t) ~.
\end{eqnarray}
\noindent
where now $x_{1}$, $x_{2}$ and $x_{3}$ denote the fractions of honest, susceptible and tax evader individuals, respectively. In addition, we also have the normalization condition
\begin{equation} \label{eq8}
x_1(t) + x_2(t) + x_3(t) = 1 ~,  
\end{equation}
\noindent
that is valid at each time step $t$. For $\epsilon=0$ we recover the model strudied in \cite{rafael}.

\begin{figure}[t]
\begin{center}
\vspace{6mm}
\includegraphics[width=0.45\textwidth,angle=0]{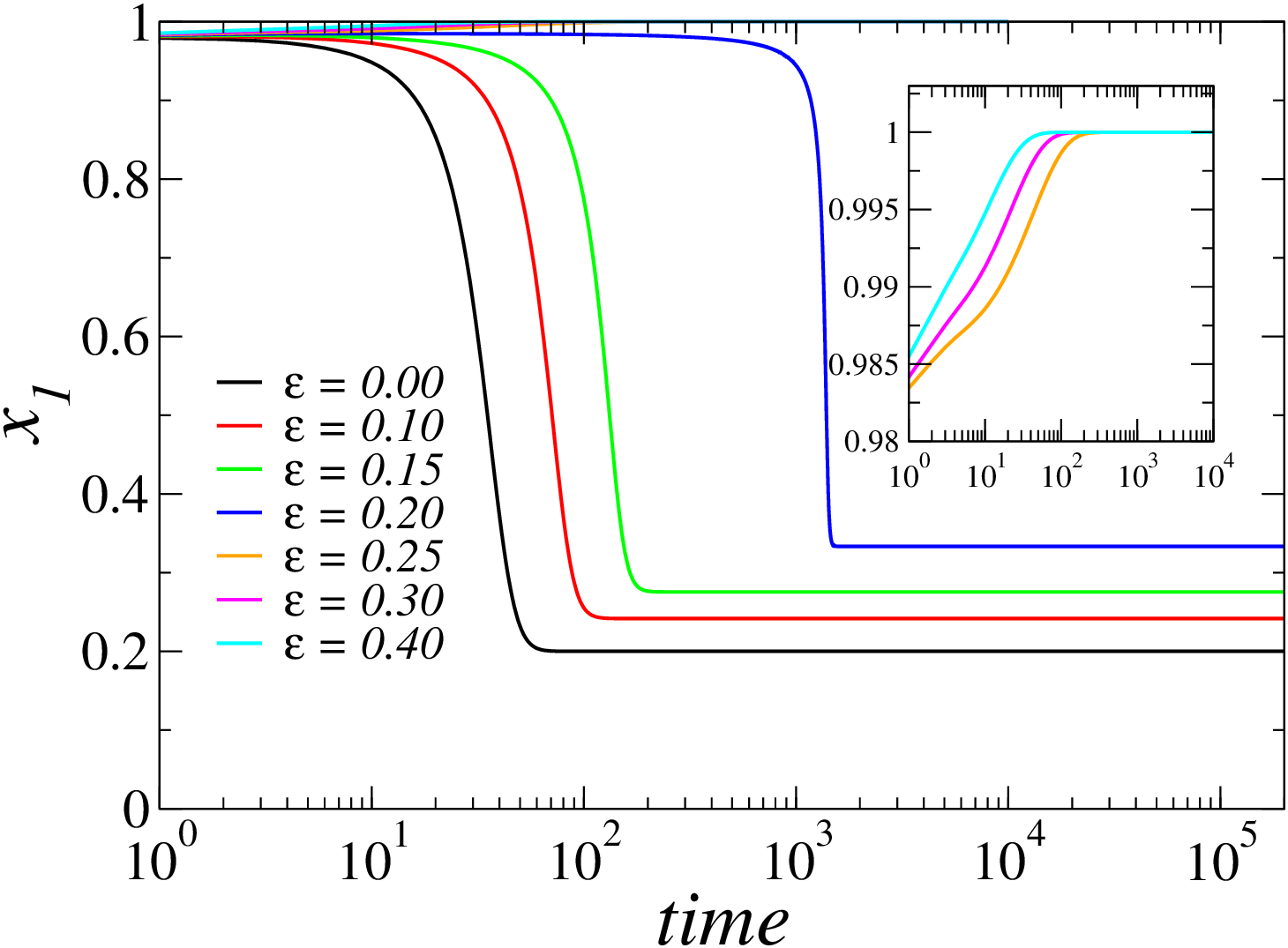}
\hspace{0.3cm}
\includegraphics[width=0.45\textwidth,angle=0]{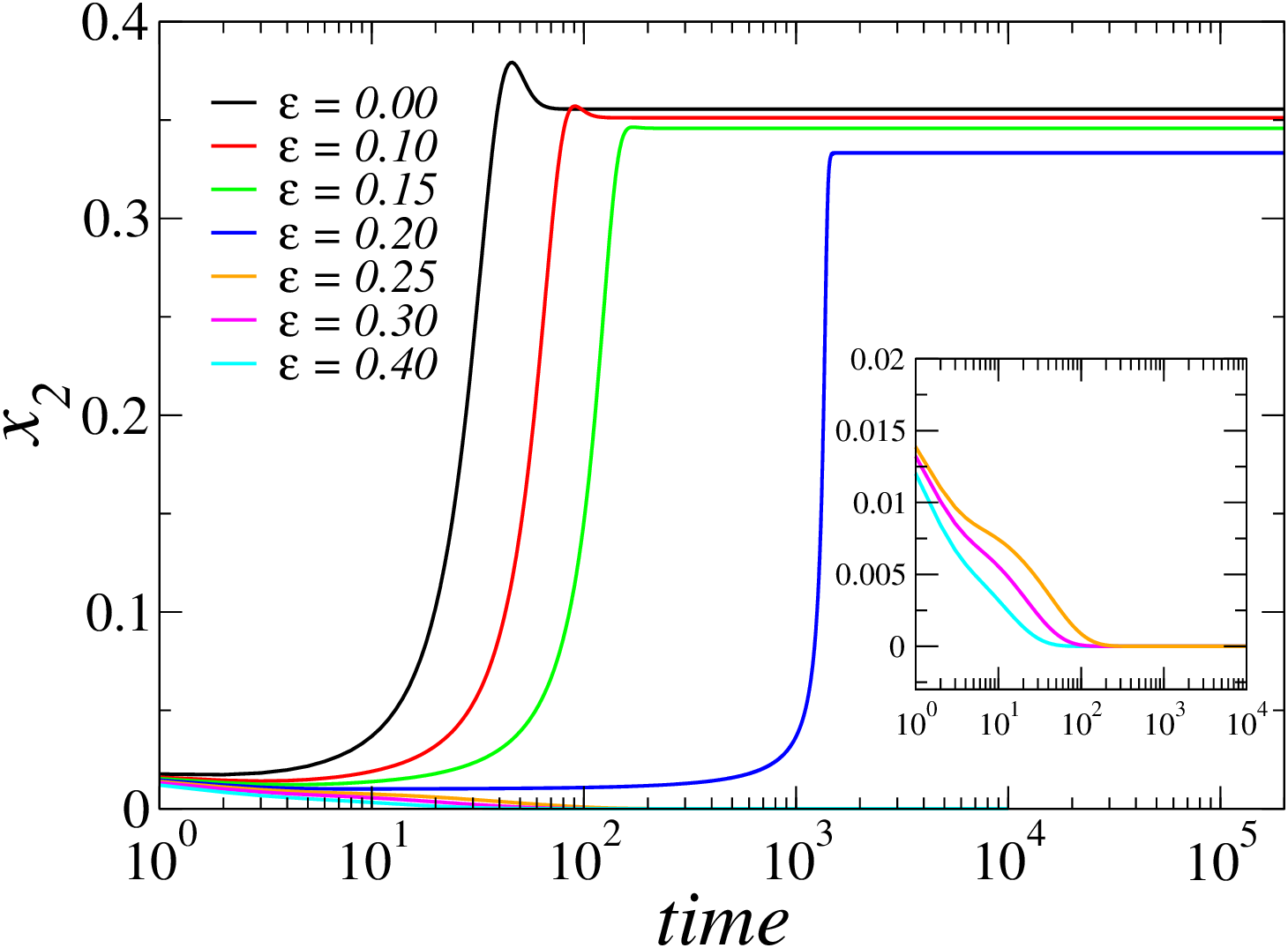}
\\
\vspace{1.0cm}
\includegraphics[width=0.45\textwidth,angle=0]{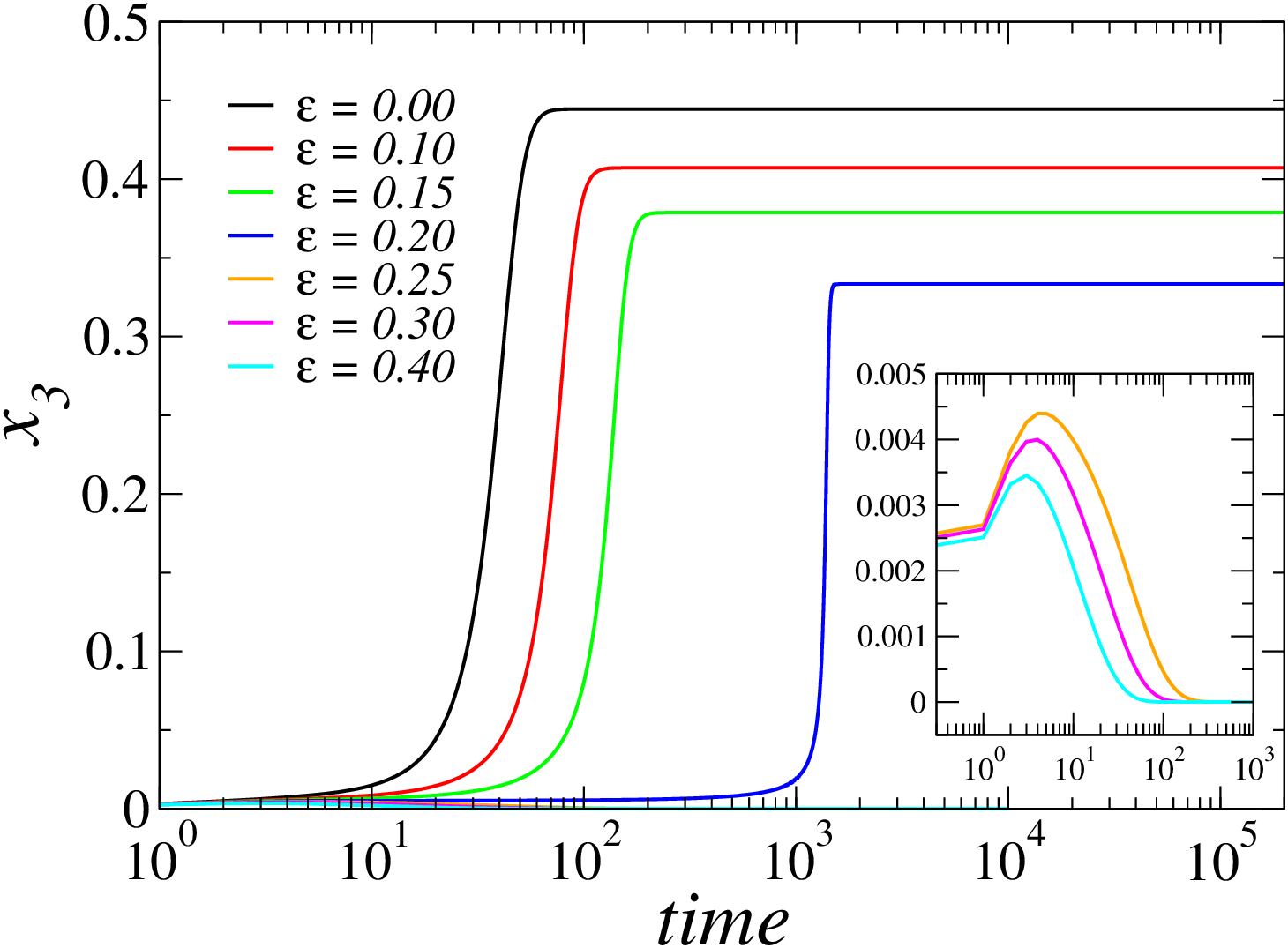}
\end{center}
\caption{(Color online) Time evolution of the three densities of agents $x_{1}$, $x_{2}$ and $x_{3}$, obtained from the numerical integration of Eqs. \eqref{eq5} - \eqref{eq7} for typical values of $\epsilon$. The fixed parameters are $\alpha=0.2$, $\beta=0.1$, $\delta=0.3$ and $\lambda=0.8$, and we vary the value of $\epsilon$. The insets show the relaxation of the densities to the absorbing state, where $x_1=1$ and $x_2=x_3=0$.}
\label{fig1}
\end{figure}

One can start analyzing the time evolution of the three classes of individuals. For this purpose, we numerically integrated  Eqs. (\ref{eq5}), (\ref{eq6}) and (\ref{eq7}). As initial conditions, we considered $x_{1}(0)=0.98$, $x_{2}(0)=0.02$ and $x_{3}(0)=0.00$. In Fig. \ref{fig1} we exhibit results for the time evolution of the fractions $x_1, x_2$ and $x_3$. As $\epsilon$ is the new parameter of the model, we fixed $\alpha=0.2$, $\beta=0.1$, $\delta=0.3$ and $\lambda=0.8$ and plot graphics for typical values of $\epsilon$. One observe that for increasing values of $\epsilon$ the final (steady state) fractions of susceptibles and evaders decrease, and for some cases they reach the steady state values $x_2=x_3=0$. On the other hand, the stationary fraction of honest tax payers increases for increasing values of $\epsilon$, and for sufficient large $\epsilon$ we have $x_1=1$. This suggests the existence of a absorbing state, since if all the individuals in the populations are honests, the system remains frozen (see the rules given by Eqs. \eqref{eq1} - \eqref{eq4}). In the next we will obtain such absorbing state analytically.

In the steady states ($t\to\infty$) we have that the three time derivatives of Eqs. \eqref{eq5} - \eqref{eq7} are zero. In such a case, Eq. \eqref{eq6} gives us
\begin{equation}\label{eq9}
x_2 = \frac{\lambda\,x_1\,x_3}{\alpha + \epsilon\,x_1}
\end{equation}

In addition, from Eq. \eqref{eq7} we have 
\begin{equation}\label{eq10}
x_2 = \frac{\beta+\delta\,x_1}{\alpha}\,x_3
\end{equation}

From Eqs. \eqref{eq9} and \eqref{eq10}, we have two solutions. The first one is given by $x_3=0$, which means that there is no evaders in the long-time limit. If this solutions is valid, from Eq. \eqref{eq9} or \eqref{eq10} we have that $x_2=0$. Thus, from the normalization condition Eq. \eqref{eq8} we have $x_1=1$. This solution $\{x_1,x_2,x_3\} = \{1,0,0\}$ represents the above-mentioned absorbing phase, since there are only honest tax payers in the population, and the dynamics become frozen. This same absorbing phase was observed in the case $\epsilon=0$ \cite{rafael}.

On the other hand, the second solution of Eqs. \eqref{eq9} and \eqref{eq10} gives us a second order polynomial for $x_1$ of the type $A\,x_1^{2} + B\,x_{1} + C = 0$, where
\begin{eqnarray} \label{eq11}
A & = & \delta\epsilon  \\ \label{eq12}
B & = & \beta\epsilon + \alpha(\delta-\lambda)  \\  \label{eq13}
C & = & \alpha\beta
\end{eqnarray}

For the limiting case $\epsilon=0$ we have $A=0$ and we recover the simple solution obtained in \cite{rafael}, namely $x_1=\frac{\beta}{\lambda-\delta}$. For $\epsilon\neq 0$, the second order polynomial gives us two solutions given by
\begin{equation} \label{eq14}
x_1 = \frac{[\beta\epsilon+\alpha(\delta-\lambda)]}{2\delta\epsilon}\left\{-1\pm\sqrt{1-\frac{4\delta\epsilon\alpha\beta}{[\beta\epsilon+\alpha(\delta-\lambda)]^{2}}}\right\}
\end{equation}

Numerically, we can verify that the physically acceptable solution is given by the plus signal in Eq. \eqref{eq14}. Substituting the normalization condition written as $x_2=1-x_1-x_3$ in Eq. \eqref{eq10}, we can obtain
\begin{equation} \label{eq15}
x_3 = \frac{\alpha(1-x_1)}{\alpha+\beta+\delta\,x_1}
\end{equation}
Given the solution of Eq. \eqref{eq14} with the plus signal, the stationary fraction of tax evaders $x_3$ can be obtained from Eq. \eqref{eq15}.

Looking for the two obtained solutions in Eq. \eqref{eq14}, and remembering that $x_1$ is not the order parameter (that is $x_3$), we expect the $x_1$ jumps at the critical point $\lambda=\lambda_c$ from the absorbing state solution $x_1=1$ to the solution given by Eq. \eqref{eq14}. In such a case, taking $x_1=1$ in Eq. \eqref{eq14}, we obtain the critical point,
\begin{equation} \label{eq16}
\lambda_c = (\beta+\delta)\,\left(1+\frac{\epsilon}{\alpha}\right) ~.
\end{equation}

\begin{figure}[t]
\begin{center}
\vspace{6mm}
\includegraphics[width=0.45\textwidth,angle=0]{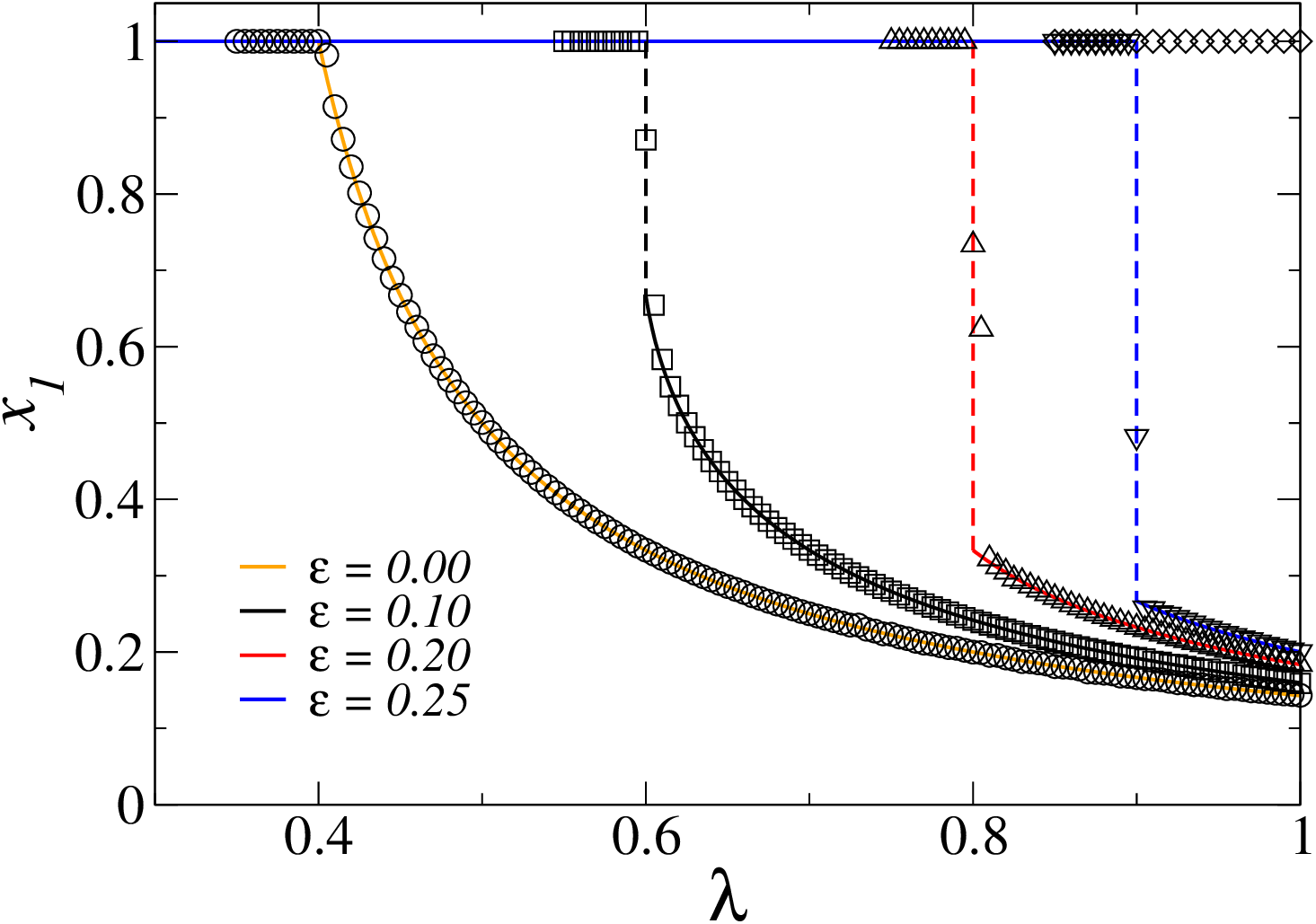}
\hspace{0.3cm}
\includegraphics[width=0.45\textwidth,angle=0]{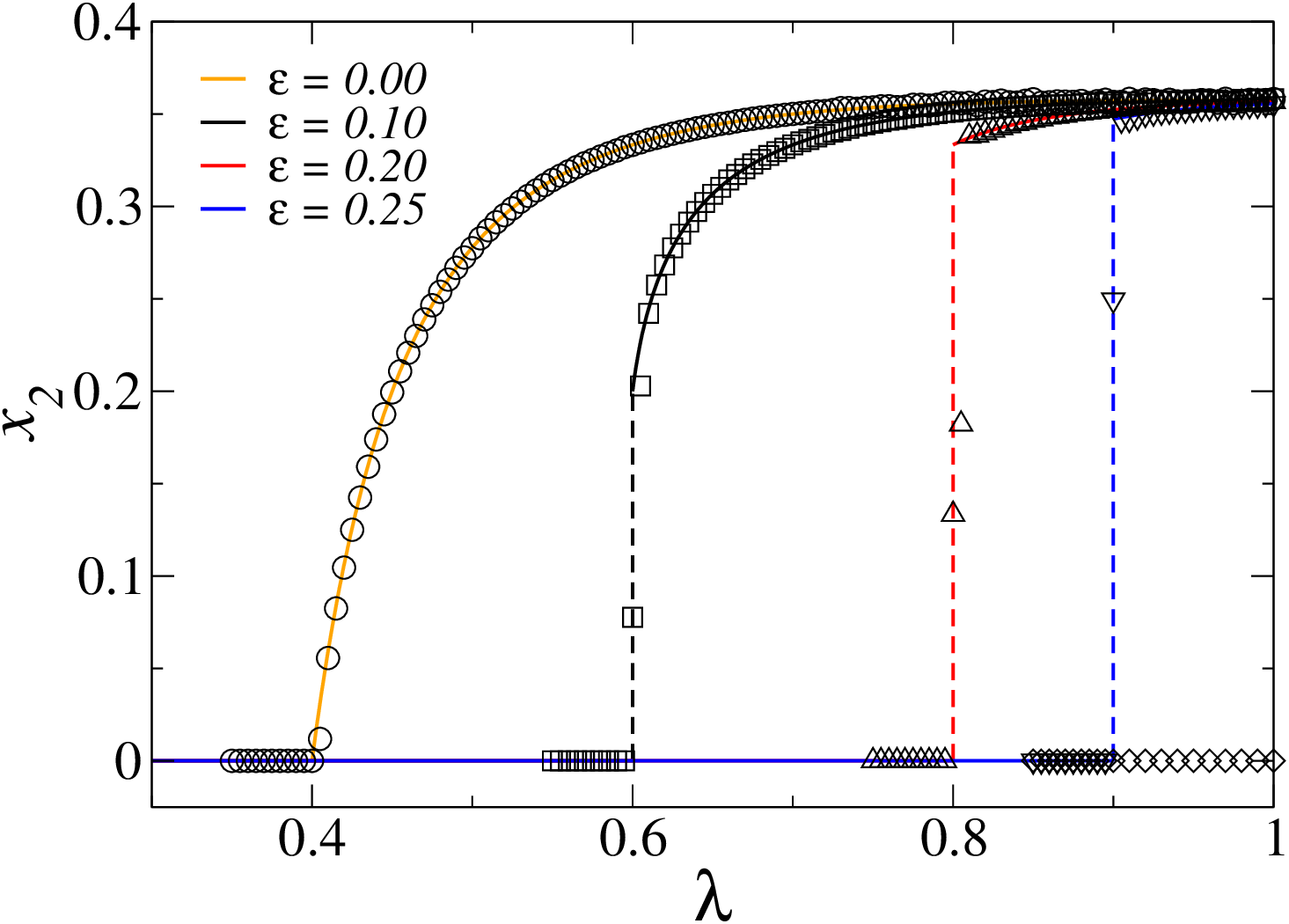}
\\
\vspace{1.0cm}
\includegraphics[width=0.45\textwidth,angle=0]{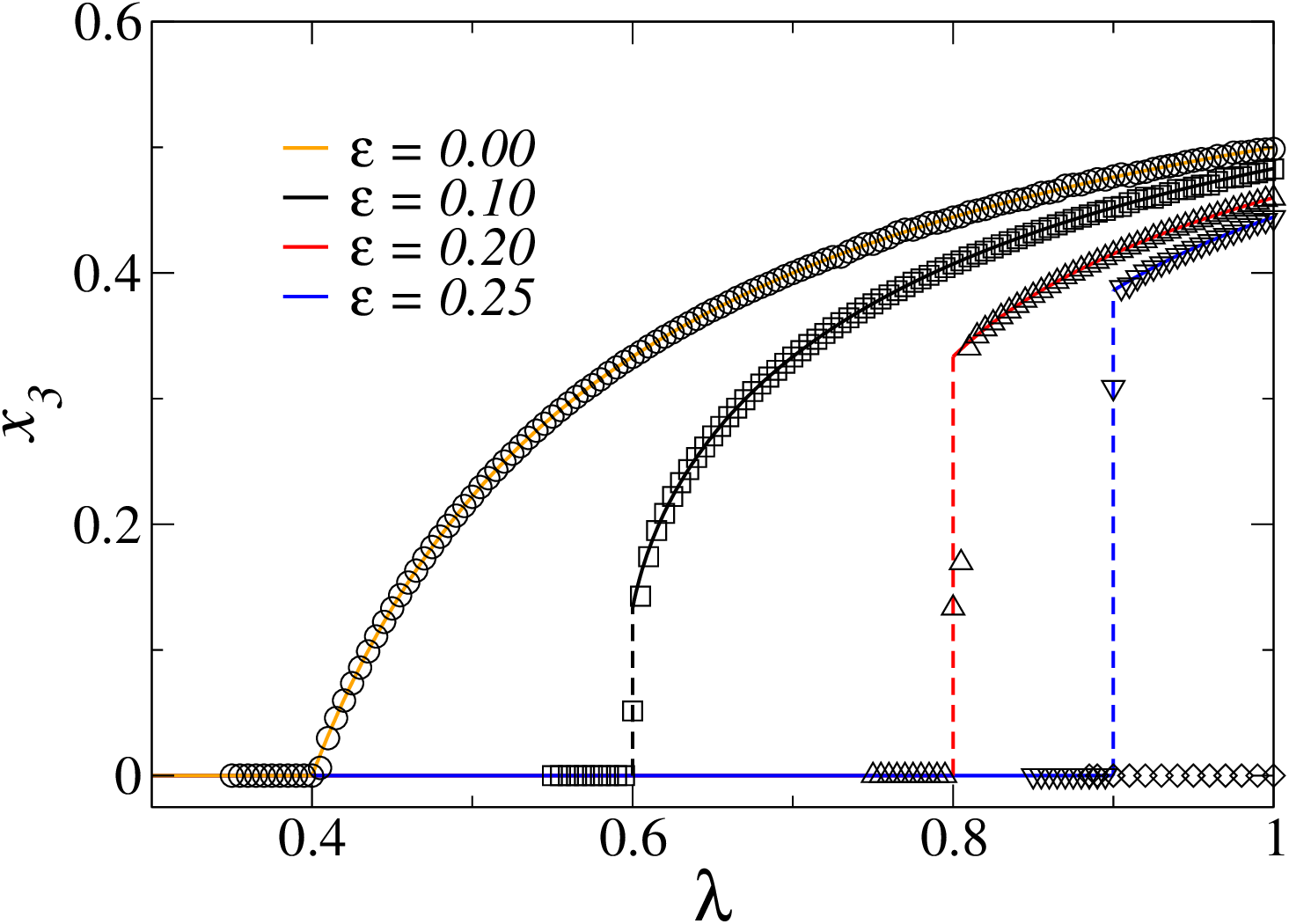}
\end{center}
\caption{(Color online) Stationary fractions of honests, susceptibles and tax evaders $x_{1}$, $x_{2}$ and $x_{3}$ as functions of $\lambda$. The full lines are the analytical results given by Eqs. \eqref{eq14}, \eqref{eq10} and \eqref{eq15}, whereas the symbols were obtained from Monte Carlo simulations with population size $N=10^{4}$, averaged over $50$ independent runs. The fixed parameters are $\alpha=0.2$, $\beta=0.1$ and $\delta=0.3$, and we considered typical values of $\epsilon$.}
\label{fig2}
\end{figure}

Notice that for the limiting case $\epsilon=0$, Eq. \eqref{eq16} recovers the result $\lambda_c=\beta+\delta$ obtained in Ref. \cite{rafael}. The behavior of Eq. \eqref{eq14} is typical of first-order phase transitions, as observed for example in opinion dynamics models \cite{biswas_pre}, ecological dynamics \cite{windus1,windus2} and quantum spin system with long-range interaction \cite{jo}. Thus, taking into account the solutions given by $x_1=1$ and Eq. \eqref{eq14}, we expect that the fraction of honests $x_1$ presents a jump at the critical points $\lambda_c$ given by Eq. \eqref{eq16}. In such a case, the order parameter $x_3$ also presents a discontinuity at $\lambda=\lambda_c$. Taking into account the two solutions for the order parameter $x_3$, namely $x_3=0$ and the other given by Eq. \eqref{eq15}, we expect to observe nonequilibrium first-order phase transitions in the model at the critical points $\lambda_c$, obtained in the terms of the models' parameters and given by Eq. \eqref{eq16}. For $\lambda>\lambda_c$ we have a phase where the three fractions $x_1, x_2$ and $x_3$ coexist. On the other hand, for $\lambda<\lambda_c$, the valid solution is given by $x_1=1, x_2=x_3=0$. In this case, we are talking about a first-order active-absorbing transition \cite{dickman,hinrichsen} that separates a phase where the tax evaders disappear of the population in the long-time limit and the population is formed only by honests, from a phase where there is a finite fraction of evaders in the long time. The susceptible agents also survive in the active phase, and they disappear in the absorbing phase. To the best of our knowledge, it is the first time that an active-absorbing nonequilibrium phase transition is observed in models of tax evasion. However, such kind of transition was observed in a wide range of systems, for example coupled opinion-epidemic dynamics \cite{jstat_andre_marcelo}, one-dimensional long-range contact processes \cite{fiore1}, Ziff-Gulari-Barshad (ZGB) model \cite{fiore2,ferreira}, granular systems \cite{neel}, opinion dynamics \cite{pre_andre_marcelo,gambaro}, naming games \cite{edgardo1,edgardo2}, symbiotic contact process \cite{soft} and majority-vote model \cite{chen}, among others.

One can also estimate the limiting case $\epsilon_c$ above which there is no phase transition anymore. Taking $\lambda_c(\epsilon_c)=1$ in Eq. \eqref{eq16}, we have
\begin{equation} \label{eq17}
\epsilon_c = \alpha\,\left(\frac{1}{\beta+\delta} - 1\right) ~. 
\end{equation}

To test the above analytical results, we performed numerical simulations of the model in a fully-connected graph. In Fig. \ref{fig2} we exhibit the stationary fractions $x_1, x_2$ and $x_3$ as functions of $\lambda$ for typical values of $\epsilon$. The fixed parameters are $\alpha=0.2$, $\beta=0.1$ and $\delta=0.3$. The full lines are the analytical results for $x_1, x_2$ and $x_3$ given by Eqs. \eqref{eq14}, \eqref{eq10} and \eqref{eq15}, respectively, and the symbols were obtained from simulations with population size $N=10^{4}$. For the simulations, the distance between two data points is $\Delta\lambda = 0.005$, except for the case where $\epsilon=0.35$, for better visualization (see the black diamonds in Fig. \ref{fig2}). The analytical and numerical results are in excellent agreement. The limiting case $\epsilon=0$ was obtained from the expressions of Ref. \cite{rafael}, and they are exhibited only for comparison with the cases where $\epsilon\neq 0$. As previous discussed, in the absence of the hesitance represented by the probability $\epsilon$, the model undergoes a continuous phase transition. On the other hand, in the presence of the interaction among honests and susceptibles, the transition is of first-order type. Notice that, for the parameters considered in Fig. \ref{fig2}, we have from Eq. \eqref{eq17} $\epsilon_c=0.3$. Thus, we expect that for $\epsilon>0.3$ the system does not undergoes a phase transition, which is exact we observe in Fig. \ref{fig2} for $\epsilon=0.35$ (see the black diamonds).

\begin{figure}[t]
\begin{center}
\vspace{6mm}
\includegraphics[width=0.6\textwidth,angle=0]{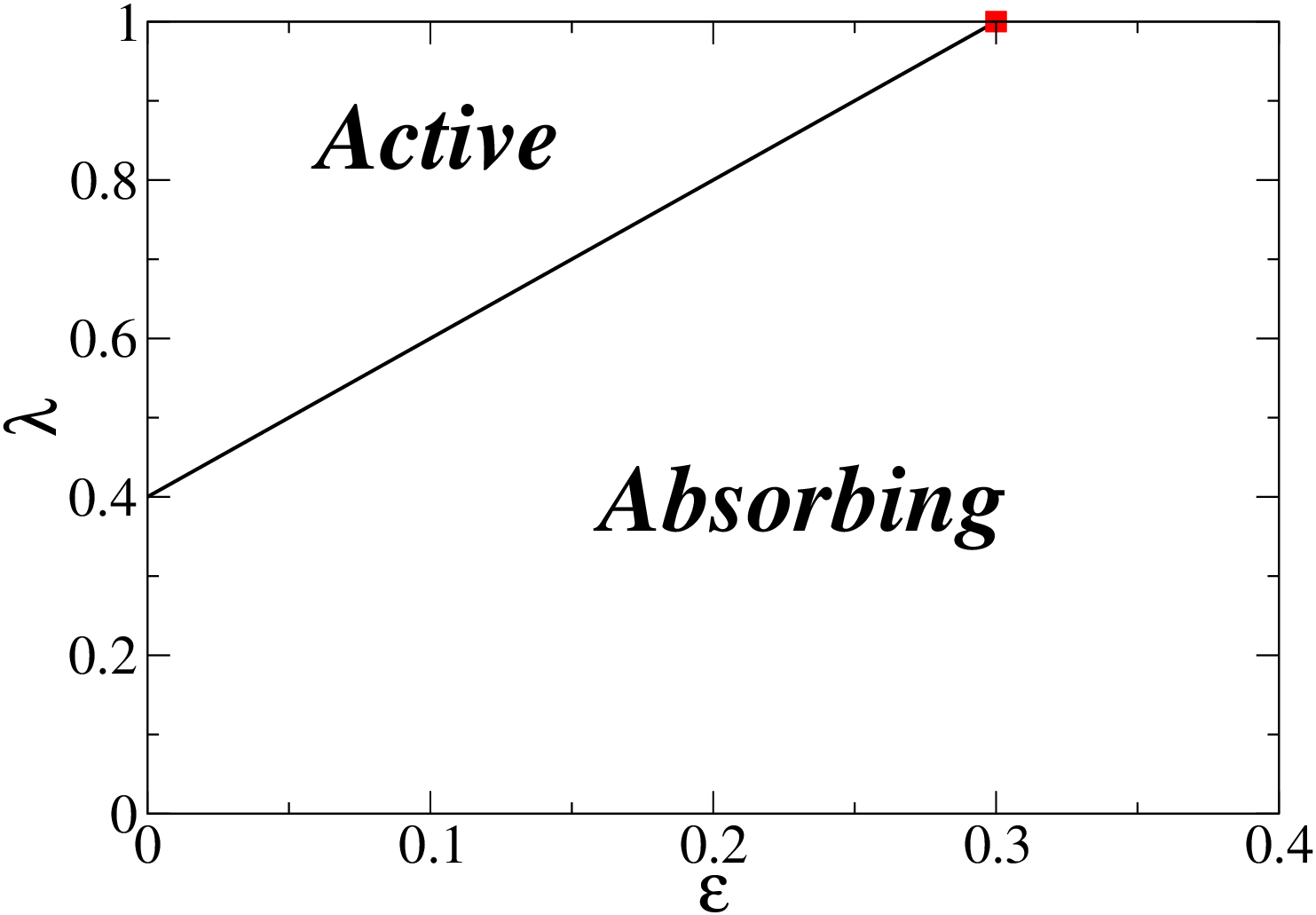}
\end{center}
\caption{(Color online) Phase diagram of the model in the plane $\lambda$ vs $\epsilon$. The full (black) line is given by Eq. \eqref{eq16}, and the (red) square represents the critical point $\epsilon_c$ given by Eq. \eqref{eq17}, which represents the limiting case above which there is no phase transition anymore. The fixed parameters are $\alpha=0.2$, $\beta=0.1$ and $\delta=0.3$.}
\label{fig3}
\end{figure}

To summarize the results, we exhibit in Fig. \ref{fig3} the phase diagram of the model in the plane $\lambda$ vs $\epsilon$ for fixed parameters $\alpha=0.2$, $\beta=0.1$ and $\delta=0.3$, separating the Absorbing and Active phases. The full (black) line is given by Eq. \eqref{eq16}, and the (red) square represents the critical point $\epsilon_c$ given by Eq. \eqref{eq17}, which represents the above-discussed limiting case.


\section{Concluding remarks}   

\qquad In this work, we have studied the dynamics of tax evasion based on a contagion model. The model consists of three distinct compartments, namely honest tax payers, tax evaders and susceptibles, an intermediate compartment between honests and evaders. We considered a fully-connected population, and we introduce a new transition probability $\epsilon$ in the model of Ref. \cite{rafael}, that can be related to the hesitance of susceptible individuals to become evaders, due to social influence of honest tax payers.

In such a case, we derive the mean-field master equations that allow us to analyze the dynamics and the steady-state properties of the model. For the limiting case $\epsilon=0$ the model undergoes a continuous nonequilibrium phase transition from an active phase, where the three compartments coexist in the population, to an absborbing phase where all individuals are honests \cite{rafael}. For the present case, our analytical and numerical results show that the introduction of only a new parameter, $\epsilon$, leads the system to undergoes nonequilibrium first-order phase transitions at critical points $\lambda_c$ that depend on the other four parameters of the model ($\alpha, \beta, \delta$ and $\epsilon$). In other words, the presence of the above-mentioned hesitance turns the continuous phase transition into a discontinuous one. Since it is not usual to observe even continuous phase transitions in models of tax evasion, the present model presents theoretical interest for the Statistical Physics and Complex Systems community.

We built a phase diagram of the model. Such phase diagram shows the existence of two states, namely the Absorbing state and the Active state. The nature of such phases are completely distinct: whereas in the Active phase we have the coexistence of the 3 populations ($X_1$, $X_2$ and $X_3$), in the Absorbing phase only the honest tax payers ($X_1$) survive in the stationary states, leading to a frozen state that is characteristic of an Absorbing state \cite{dickman}. In addition, the transition between those two states can be of continuous type (for $\epsilon=0$) or first-order type (for $\epsilon>0$).

As future extensions, it can be considered the inclusion of heterogeneities in the population, as the presence of special agents like contrarians individuals \cite{gambaro,galam_cont,gordon}, zealots or sttuborn individuals  \cite{laguna1,laguna2,galam1,galam2,mobilia} and opinion leaders \cite{boccara,zhu,liu}. The presence of distinct kinds of noises can also be considered \cite{lalama1,lalama2,pre2013}, among others. The properties of this model in various lattices and networks would also be interesting to analyze.


\section*{Acknowledgments}

The author acknowledges financial support from the Brazilian scientific funding agencies Conselho Nacional de Desenvolvimento Cient\'ifico e Tecnol\'ogico (CNPq, grant number 310893/2020-8) and Funda\c{c}\~ao Carlos Chagas Filho de Amparo \`a Pesquisa do Estado do Rio de Janeiro (FAPERJ, grant number 203.217/2017).

\end{document}